\documentclass[twocolumn,amsmath,amssymb,floatfix,prl,shownopacs,footinbib,superscriptaddress]{revtex4-1}
\usepackage{amsmath,amssymb,bm,graphicx,url,epsf,color}
\usepackage[british]{babel}
\usepackage{wasysym}
\usepackage{MnSymbol}

\newcommand{\Tohm}{T_{\Omega}}
\newcommand{\Jin}{J_Q}
\newcommand{\Jeph}{J_Q^{e-ph}}
\newcommand{\Je}{J_Q^{e}}
\newcommand{\QPCun}{\mathrm{QPC}_1}
\newcommand{\QPCdeux}{\mathrm{QPC}_2}

\begin{document}

\title{Quantum limit of heat flow across a single electronic channel}

\author{S. Jezouin}
\thanks{These authors contributed equally to this work.}
\affiliation{CNRS, Laboratoire de Photonique et de Nanostructures
(LPN), 91460 Marcoussis, France}
\author{F.D. Parmentier}
\thanks{These authors contributed equally to this work.}
\affiliation{CNRS, Laboratoire de Photonique et de Nanostructures
(LPN), 91460 Marcoussis, France}
\author{A. Anthore}
\email[e-mail: ]{anne.anthore@lpn.cnrs.fr}
\affiliation{CNRS, Laboratoire de Photonique et de Nanostructures
(LPN), 91460 Marcoussis, France}
\affiliation{Univ Paris Diderot, Sorbonne Paris Cit\'e, LPN, 91460 Marcoussis, France}
\author{U. Gennser}
\affiliation{CNRS, Laboratoire de Photonique et de Nanostructures
(LPN), 91460 Marcoussis, France}
\author{A. Cavanna}
\affiliation{CNRS, Laboratoire de Photonique et de Nanostructures
(LPN), 91460 Marcoussis, France}
\author{Y. Jin}
\affiliation{CNRS, Laboratoire de Photonique et de Nanostructures
(LPN), 91460 Marcoussis, France}
\author{F. Pierre\thanks{frederic.pierre@lpn.cnrs.fr}}
\email[e-mail: ]{frederic.pierre@lpn.cnrs.fr}
\affiliation{CNRS, Laboratoire de Photonique et de Nanostructures
(LPN), 91460 Marcoussis, France}

\begin{abstract}
Quantum physics predicts that there is a fundamental maximum heat conductance across a single transport channel, and that this thermal conductance quantum $G_Q$ is universal, independent of the type of particles carrying the heat. Such universality, combined with the relationship between heat and information, signals a general limit on information transfer. We report on the quantitative measurement of the quantum limited heat flow for Fermi particles across a single electronic channel, using noise thermometry. The demonstrated agreement with the predicted $G_Q$ establishes experimentally this basic building block of quantum thermal transport. The achieved accuracy of below 10\% opens access to many experiments involving the quantum manipulation of heat.
\end{abstract}

\maketitle

\begin{figure}[!htb]
\renewcommand{\figurename}{\textbf{Figure}}
\renewcommand{\thefigure}{\textbf{\arabic{figure}}}
\centering\includegraphics[width=1\columnwidth]{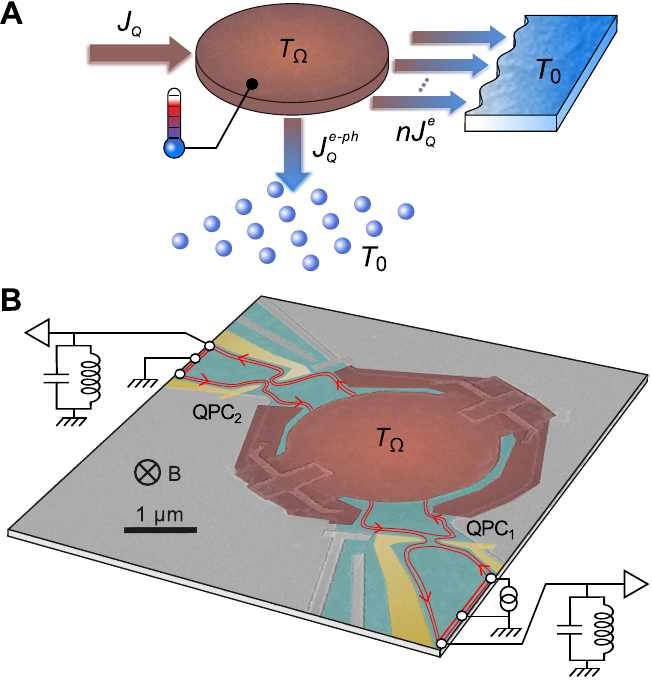}
\caption{
\textbf{Experimental principle and practical implementation.} \textbf{(A)} Principle of the experiment: electrons in a small metal plate (brown disk) are heated up to $\Tohm$ by the injected Joule power $\Jin$. The large arrows symbolize injected power ($\Jin$) and outgoing heat flows ($n\Je$, $\Jeph$). \textbf{(B)} False-colors SEM picture of the measured sample. The Ga(Al)As 2D electron gas is highlighted in light blue, the QPC metal gates in yellow and the micron-sized metallic ohmic contact in brown. The light gray metal gates are polarized with a strong negative gate voltage and are not used in the experiment. Two copropagating edge channels (shown out of $\nu=3$ or $\nu=4$) have their propagation direction indicated by red arrows. $\QPCun$ is here set to fully transmit a single channel ($n_1=1$) and $\QPCdeux$ two channels ($n_2=2$), corresponding to a total number of open electronic channels $n=n_1+n_2=3$. The experimental apparatus is shown as a simplified diagram. It includes two $L-C$ tanks used to perform the noise thermometry measurements around 700 kHz. The Joule power $\Jin$ is injected on the micron-sized metallic electrode from the DC polarization current partly transmitted through $\QPCun$.
}
\label{fig-sample}
\end{figure}

The transport of electricity and heat in reduced dimensions and at low temperatures is subject to the laws of quantum physics. The Landauer formulation of this problem \cite{Landauer1975,Anderson1980,Buttiker1986} introduces the concept of transport channels: a quantum conductor is described as a particle waveguide, and the channels can be viewed as the quantized transverse modes. Quantum physics sets a fundamental limit to the maximum electrical conduction across a single electronic channel. The electrical conductance quantum $G_e=e^2/h$, where $e$ is the unit charge and $h$ is the Planck constant, was initially revealed in ballistic 1D constrictions \cite{VanWees1988,Wharam1988}. However, different values of the maximum electrical conductance are observed for different types of charge carrying particles. In contrast, for heat conduction the equivalent thermal conductance quantum $G_Q=\pi^2k_B^2T/3h\simeq (1~\mathrm{pW}/\mathrm{K}^{2})T$ (which sets the maximum thermal conduction across a single transport channel, $k_B$ being the Boltzmann constant, $T$ the temperature) is predicted to be independent of the heat carrier statistics, from bosons to fermions including the intermediate `anyons' \cite{Pendry1983,Maynard1985,Sivan1986,Butcher1990,Kane1996,Kane1997,Greiner1997,Rego1999,Krive1999,Blencowe2000,Schmidt2004}. In electronic channels, which carry both an electrical and thermal current, the predicted ratio $(\pi^2k_B^2/3e^2)T$ between $G_Q$ and $G_e$ verifies and extends the Wiedemann-Franz relation down to a single channel \cite{Sivan1986,Butcher1990}. In general, the universality of $G_Q$, together with the deep relationship between heat, entropy and information \cite{Shannon1948}, points to a quantum limit on the flow of information through any individual channel \cite{Pendry1983,Blencowe2000}.

The thermal conductance quantum has been measured for bosons, in systems with as few as 16 phonon channels \cite{Schwab2000,Yung2002}, and probed at the single photon channel level \cite{Meschke2006,Timofeev2009}. For fermions, heat conduction was shown to be proportional to the number of ballistic electrical channels \cite{Molenkamp1992,Chiatti2006}. In \cite{Molenkamp1992} the data were found compatible, within an order of magnitude estimate, to the predicted thermal conductance quantum, whereas \cite{Chiatti2006} demonstrated more clearly the quantization of thermal transport, but $G_Q$ was not accessible by construction of the experiment.

\begin{figure}[!htb]
\renewcommand{\figurename}{\textbf{Figure}}
\renewcommand{\thefigure}{\textbf{\arabic{figure}}}
\centering\includegraphics[width=1\columnwidth]{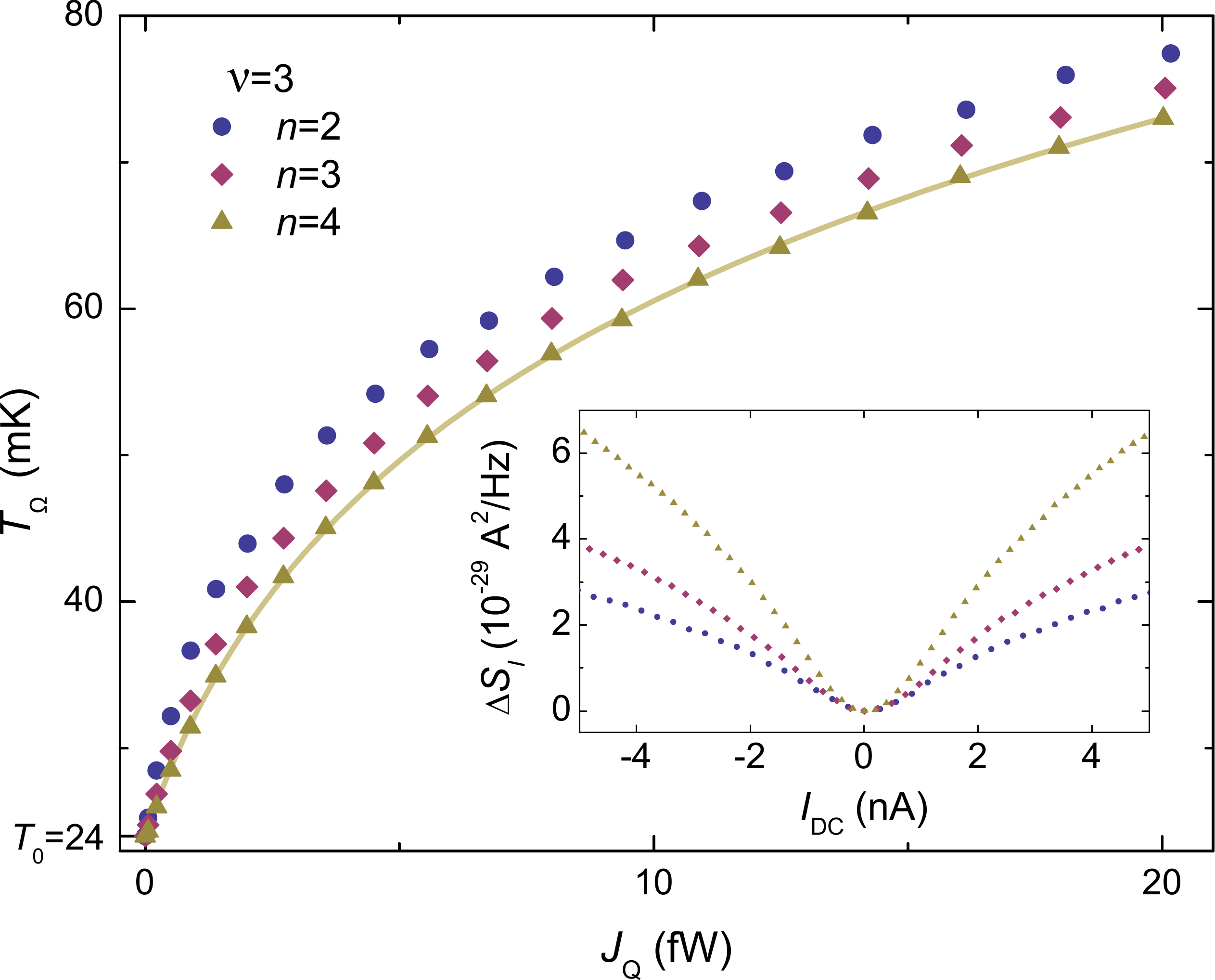}
\caption{
\textbf{Noise thermometry measurements versus injected power.} The electronic temperature $\Tohm$ in the micron-sized ohmic contact is plotted as a function of the injected power $\Jin$ at $\nu=3$ and for a base electronic temperature $T_0=24$ mK. The symbols $\bullet$, $\blacklozenge$ and $\blacktriangle$ correspond, respectively, to $n=2$, 3 and 4 open electronic channels. The continuous line is a fit of the data for $n=4$ open channels including the heat transfer to phonons (see text). The size of the symbols is indicative of the experimental error bars. Inset: measured excess noise spectral density versus applied DC current. Note that in the main panel, data at opposite DC currents, which are equal at our experimental accuracy, are averaged to improve the signal to noise ratio.
}
\label{fig-noise}
\end{figure}

\begin{figure*}[!bt]
\renewcommand{\figurename}{\textbf{Figure}}
\renewcommand{\thefigure}{\textbf{\arabic{figure}}}
\centering\includegraphics [width=1\textwidth]{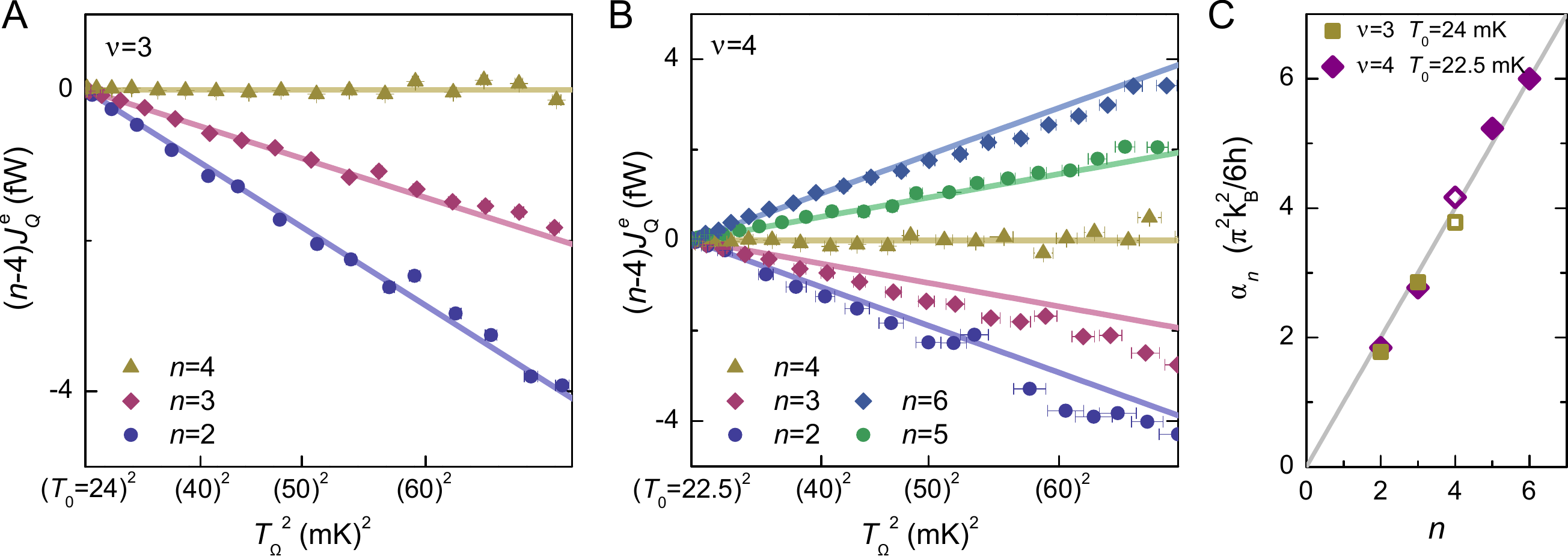}
\caption{
\textbf{Heat flow across ballistic electronic channels.} \textbf{(A),(B)} The symbols display the heat current across $|n-4|$ electronic channels, with a positive (negative) sign for $n>4$ ($n<4$), as a function of the squared temperature $\Tohm^2$ of the micron-sized ohmic contact. The data in \textbf{(A)} (\textbf{(B)}) were measured at $\nu=3$ ($\nu=4$) at a base temperature $T_0=24$ mK ($T_0=22.5$ mK) for $n=2$, 3 and 4 ($n=2$, 3, 4, 5, 6), respectively from bottom to top. The continuous lines are the theoretical predictions for the quantum limited heat flow.
\textbf{(C)} Extracted electronic heat current factor $\alpha_{n}\equiv n\Je/(\Tohm^2-T_0^2)$ in units of $\pi^2k_B^2/6h$ (symbols) versus the number $n$ of electronic channels. It is obtained from the fitted slopes $\alpha_{n-4}'$ of the data in \textbf{(A)} and \textbf{(B)} added to the separately extracted value of $\alpha_{4}$ (see text and Fig. 2): $\alpha_{n}= \alpha_{n-4}'+\alpha_{4}$, with $\alpha_{4}=3.8$ ($\alpha_{4}=4.2$) at $\nu=3$ ($\nu=4$) shown distinctly as open symbols. The predictions for the quantum limit of heat flow fall on the continuous line $y=x$.
}
\label{fig-thermcond}
\end{figure*}

We have measured the quantum limited heat flow across a single electronic channel using the conceptually simple approach depicted in Fig. 1A. A micron-sized metal plate is electrically connected by an adjustable number $n$ of ballistic quantum channels to a cold bath at temperature $T_0$. Electrons in the small plate are heated up with a well-known Joule power ($\Jin$ in Fig. 1A), and the resulting increased electronic temperature $\Tohm$ is measured by direct thermometry based on noise measurements. Heat balance implies that the injected Joule power is compensated by the overall outgoing heat current:
\begin{equation}
\Jin=n\Je(\Tohm,T_0)+\Jeph(\Tohm,T_0),
\label{eq-heatbalance}
\end{equation}
where $n\Je(\Tohm,T_0)$ is the electronic heat flow across the $n$ ballistic quantum channels. The flow $\Jeph(\Tohm,T_0)$ is an additional contribution, here attributed to the transfer of heat from the hot electrons toward the cold phonon bath in the plate (and thus independent of $n$). The heat flow across a single ballistic electronic channel is then directly given by how much $\Jin$ is increased to keep $\Tohm$ constant when one additional electronic channel is opened. The quantum limited heat flow for a single electronic channel connecting two heat baths at $\Tohm$ and $T_0$ reads \cite{Sivan1986,Butcher1990}:
\begin{equation}
\Je(\Tohm,T_0)=\frac{\pi^2 k_B^2}{6h}(\Tohm^2-T_0^2).
\label{eq-JQe}
\end{equation}
The quadratic temperature dependence reflects the fact that the temperature sets both the average energy of electronic excitations, as well as their number, the latter being proportional to the energy bandwidth.

The actual sample, displayed in Fig. 1B, was cooled down to $T_0 \approx 20$ mK in our experiment. The noise thermometry was performed with ultra-sensitive cryogenic electronics based on a homegrown high electron mobility transistor \cite{Liang2012}.

The ballistic quantum channels are formed in a high mobility Ga(Al)As two-dimensional electron gas by the field-effect tuning of two quantum point contacts (QPC), labeled $\QPCun$ and $\QPCdeux$ in Fig. 1B. The potentials applied to the metal split gates (colorized yellow in Fig. 1B) are set to fully transmit $n$ electronic channels, in which case the QPC electrical conductances display clear plateaus (see supplementary Fig. S5 in \cite{si}). The measured conductances $n_1e^2/h$ and $n_2e^2/h$ of, respectively, $\QPCun$ and $\QPCdeux$, correspond to $n=n_1+n_2$. Note that for fully transmitted channels, the electron-photon coupling observed in the same sample for the case of partially transmitted channels \cite{Jezouin2013} vanishes.

The heated-up metal plate, colorized brown in Fig. 1B, is a micron-sized ohmic contact \cite{si}, which is electrically connected to cold electrodes located further away exclusively through the two QPCs. In order to approach the quantum limit of heat flow per channel, the electrical connection between the plate and the 2D electron gas located 94 nm below the surface must have a negligibly low resistance compared to $h/e^2$. Moreover, the heated-up electrons must dwell in the plate for a time longer than the electron-electron energy exchange time, in order to relax toward a quasi-equilibrium situation characterized by a hot Fermi distribution at $\Tohm$. These two conditions set the minimum size of the ohmic contact \cite{si}; on the other hand, the ohmic contact must be small enough to minimize the heat transfer towards phonons, which is proportional to volume. The sample was optimized to fulfill these antagonistic requirements, achieving a negligibly small contact resistance, a typical dwell time in the $10~\mu$s range, and a dominating electronic heat flow for $\Tohm \lesssim 70$ mK.

The sample was subjected to a strong perpendicular magnetic field in order to enter the integer quantum Hall effect (QHE) regime, at filling factors $\nu=3$ or $\nu=4$. In this regime, the current flows along the sample edges in so-called edge channels with a unique propagation direction (continuous red lines with arrows in Fig. 1B). One motivation for performing the experiment in this regime is the spatial separation between incoming and outgoing edge channels away from the QPC, which enables using the large metal electrodes located further away as ideal cold reservoirs. Furthermore, it is easier to tune the QPCs to a discrete set of fully open channels \cite{si}, and the spin degeneracy is broken so that the electronic channels can be opened one at a time. Finally, the QHE regime allows for a simple implementation of the noise thermometry.

The injected Joule power $\Jin$ was generated with an applied DC current (Fig. 1B) partly transmitted across the $n_1$ ballistic channels of $\QPCun$ into the plate \cite{si}.

The resulting increase $\Tohm-T_0$ of the electronic temperature in the plate was determined from the increase $\Delta S_I$ in the measured spectral density of the current noise along the outgoing edge channels \cite{Buttiker1990,Blanter2000chaotic,si}:
\begin{equation}
\Delta S_I = 2k_B (\Tohm-T_0) \frac{G_e}{1/n_1+1/n_2}.
\label{eq-ExcessNoise}
\end{equation}

The raw measurements of excess current noise $\Delta S_I$ versus applied DC current for $n\in\{2,3,4\}$ open electronic channels at $\nu=3$ are shown as symbols in the inset of Fig. 2. Here $n=2$ corresponds to $(n_1,n_2)=(1,1)$, $n=3$ is the average over the two equivalent configurations $(1,2)$ and $(2,1)$ (see supplementary Fig. S6 in \cite{si}), and $n=4$ corresponds to $(2,2)$. The displayed data are measured on the top left electrode, behind $\QPCdeux$; the same excess noise, within $2\%$, was measured with another amplification chain on the bottom right electrode, behind $\QPCun$ \cite{si}.

The main panel of Fig. 2 shows these data recast as the measured electronic temperature in the micron-sized plate $\Tohm$ versus the injected Joule power $\Jin$.
The base temperature $T_0=24$ mK was obtained separately, from a noise thermometry performed at $\nu=3$ during the same experimental run \cite{si}.
At fixed $\Tohm$, the distinct increase in $\Jin$ as the number $n$ of ballistic electronic channels is incremented by one directly corresponds to the heat flow across an individual electronic channel. We focus here on the low injected power regime $\Jin\lesssim 20$ fW, where the electronic heat flow $n\Je$ is the most important contribution at $n=4$ \cite{si}.

We now separate the electronic heat flow $n\Je$ from the additional, a priori unidentified, contribution $\Jeph$, which is independent of the number $n$ of open electronic channels. For this purpose, we use as a reference the $n=4$ data corresponding to the most robust QPC plateaus $n_1=n_2=2$.

First, we show that important information can already be extracted at $n=4$ within a model-dependent approach. The $n=4$ reference data $\Jin=\Jeph+4\Je$ are fitted using the standard expression for electron-phonon cooling in diffusive metals \cite{Wellstood1994,Pierre2003,Giazotto2006} $\Jeph(\Tohm,T_0)=\Sigma\Omega(\Tohm^5-T_0^5)$ with $\Sigma\Omega$ a free parameter; for $4\Je$ we used the predicted power law (see Eq. 2), $4\Je=\alpha_4 (\Tohm^2-T_0^2)$ with $\alpha_4$ a free parameter. The fit is shown as a continuous yellow line in Fig. 2. The electron-phonon coupling parameter $\Sigma\Omega=5.5~\mathrm{nW}/\mathrm{K}^5$ extracted from the fit is a very typical value for similar metals \cite{Wellstood1994,Pierre2003,Giazotto2006} given the micron-sized ohmic contact volume $\Omega\sim 2~\mathrm{\mu m}^3$. The extracted electronic heat flow $4\Je$ is found to be $3.8 (\pi^2 k_B^2/6h)(\Tohm^2-T_0^2)$, which is within $5~\%$ of the theoretically predicted value given by Eq. 2. The same $\Sigma\Omega$ and electronic heat current are obtained by repeating this analysis on the $n=4$ data at the higher temperature $T_0=40$ mK, whereas a relatively small difference of about $20~\%$ is seen at filling factor $\nu=4$ (see supplementary Figs. S8 and S9 in \cite{si}).

Second, following the model-free approach described earlier, we extract the amount of heat $(n-4)\Je(\Tohm,T_0)$ flowing across the additional $(n-4)$ ballistic electronic channels by subtracting from the measured $\Jin$ with $n$ open channels the $n=4$ reference signal. At $n>4$ ($n<4$), $|n-4|$ channels are opened (closed) with respect to the reference configuration. Given the extremely accurate fit described above, within experimental error bars, we choose to subtract the fit function, instead of using an arbitrary interpolation function between the measured $n=4$ data points.
The extracted variations of the electronic heat currents $(n-4)\Je$ are plotted as symbols in Fig. 3A as a function of the squared temperature $\Tohm^2$, for the data at $\nu=3$ and $T_0=24$ mK. Similar data obtained for $T_0=22.5$ mK at the different filling factor $\nu=4$ and for up to $n=6$ ballistic electronic channels (Fig. 3B) show larger scatter simply because of a lower experimental accuracy, due to the less favorable current-voltage conversion at this filling factor and a smaller acquisition time per point. We find that $(n-4)\Je$ is proportional to $\Tohm^2$, as expected from theory (Eq. 2).

We now compare the extracted electronic heat currents with the quantitative prediction Eq. 2 for the quantum limited heat flow.
The experimentally extracted $(n-4)\Je(\Tohm,T_0)$ are in good agreement with the theoretical predictions shown as continuous lines in Fig. 3A,B.
We then extract a quantitative experimental value for the quantum limited heat flow by fitting the data $(n-4)\Je(\Tohm,T_0)$ using the predicted and observed functional $\alpha_{n-4}'(\Tohm^2-T_0^2)$, with $\alpha_{n-4}'$ the only free parameter. Taking altogether the set of normalized slopes $\{\alpha_{n-4}'/(n-4)\}$ obtained within the model-free approach at $\nu=3$ and $\nu=4$, we find:
\begin{equation}
\frac{\Je(\Tohm,T_0)}{\Tohm^2-T_0^2} = (1.06 \pm 0.07)\times\frac{\pi^2 k_B^2}{6h},
\end{equation}
in agreement, at our experimental accuracy, with the theoretical prediction for the quantum limited heat flow across a single channel, and therefore with the predicted value given by the thermal conductance quantum. The displayed $7~\%$ uncertainty is the standard error on the mean value obtained from the 6 values $\{\alpha_{n-4}'/(n-4)\}$, each weighted by the corresponding number $|n-4|$ of electronic channels \cite{si}. Note that this uncertainty ignores systematic sources of error, e.g. on the calibrated gain of the amplification chain \cite{si}.
The accuracy can be improved by including the values of $\alpha_4$ obtained at $\nu=3$ and $\nu=4$ within the model-dependent approach detailed earlier. Figure 3C displays as symbols the full electronic heat current factors $\{\alpha_n\equiv \alpha_{n-4}'+\alpha_4\}$ versus $n$, with the corresponding theoretical predictions $n \times \pi^2 k_B^2/6h$ falling on the continuous line. The same statistical analysis on the 8 values of $\{\alpha_n/n\}$ yields $\Je(\Tohm,T_0)/(\Tohm^2-T_0^2)=(0.98 \pm 0.02)\times \pi^2 k_B^2/6h$ \cite{si}.

The present experiment demonstrates that the quantum limited heat flow across a single electronic channel, which sets the scale of quantum interference effects, is now attainable at a few \% accuracy level. This opens access to many studies in the emergent field of quantum heat transport \cite{Dubi2011,Giazotto2012}, such as quantum phase manipulation of heat currents.

\subsection*{Acknowledgments}
The authors gratefully acknowledge the contribution of V. Andreani to the noise measurement setup. This work was supported by the ERC (ERC-2010-StG-20091028, \#259033).

\end{document}

\title{Supplementary Materials for \\`Quantum limit of heat flow across a single electronic channel'}

\author{S. Jezouin}
\thanks{These authors contributed equally to this work.}
\affiliation{CNRS, Laboratoire de Photonique et de Nanostructures
(LPN), 91460 Marcoussis, France}
\author{F.D. Parmentier}
\thanks{These authors contributed equally to this work.}
\affiliation{CNRS, Laboratoire de Photonique et de Nanostructures
(LPN), 91460 Marcoussis, France}
\author{A. Anthore}
\email[e-mail: ]{anne.anthore@lpn.cnrs.fr}
\affiliation{CNRS, Laboratoire de Photonique et de Nanostructures
(LPN), 91460 Marcoussis, France}
\affiliation{Univ Paris Diderot, Sorbonne Paris Cit\'e, LPN, 91460 Marcoussis, France}
\author{U. Gennser}
\affiliation{CNRS, Laboratoire de Photonique et de Nanostructures
(LPN), 91460 Marcoussis, France}
\author{A. Cavanna}
\affiliation{CNRS, Laboratoire de Photonique et de Nanostructures
(LPN), 91460 Marcoussis, France}
\author{Y. Jin}
\affiliation{CNRS, Laboratoire de Photonique et de Nanostructures
(LPN), 91460 Marcoussis, France}
\author{F. Pierre\thanks{frederic.pierre@lpn.cnrs.fr}}
\email[e-mail: ]{frederic.pierre@lpn.cnrs.fr}
\affiliation{CNRS, Laboratoire de Photonique et de Nanostructures
(LPN), 91460 Marcoussis, France}

\maketitle

\subsection{Sample and measurement setup}

{\noindent\textbf{Sample.}} The sample is nanostructured by standard e-beam lithography in a 94 nm deep GaAs/Ga(Al)As two-dimensional electron gas of density $2.5 \times10^{15}~\mathrm{m}^{-2}$ and mobility $55~\mathrm{m}^2\mathrm{V}^{-1}\mathrm{s}^{-1}$. The ohmic contacts to the buried two-dimensional electron gas are obtained with a standard technique: a metallic multilayer of nickel-gold-germanium is diffused into the semiconductor by heating the sample.\\

{\noindent\textbf{Cryogenic environment.}} The measurements were performed in a dilution refrigerator. At low temperature, the electrical lines were carefully filtered and thermalized by inserting long ($0.3-1$~m) and resistive ($300~\Omega /$m) wires into 260~$\mu$m inner diameter CuNi tubes. The sample was further protected from spurious high energy photons by two shields, both at the mixing chamber temperature.\\

{\noindent\textbf{Low-frequency polarization and measurement setup.}} The sample was connected in a multi-terminal configuration with cold grounds, which is typical for samples in the quantum Hall regime (see article Fig.~1B). It was current biased with a $16~\mathrm{M}\Omega$ polarization resistor located at the $4.2~$K stage of the dilution refrigerator. Note that the current biasing is performed through a distinct experimental line and sample ohmic contact (see article Fig.~1B). Note also that such current biasing is immune to the thermoelectric voltages developing along the lines between room and base temperature (this can be checked very directly on supplementary Fig.~S4 from the very precise location of the dip near zero bias current). The electrical conductances are obtained by three point measurements, using standard lock-in techniques at frequencies below 100~Hz. The polarization/measured currents are converted on-chip into voltages by taking advantage of the well defined quantum Hall resistance to ground. \\

{\noindent\textbf{Injected Joule power in the micron-sized ohmic contact.}}
The applied DC current $I_\mathrm{DC}$ (see article Fig.~1B) is converted on-chip through the well-defined quantum Hall resistance into a DC voltage $V_\mathrm{DC}=I_\mathrm{DC}/(\nu G_e)$.  Using the standard Landauer-B\"{u}ttiker scattering formalism \cite{Buttiker1988}, we find that the Joule power dissipated into the micron-sized ohmic contact reads:
\begin{equation}
J_Q = \frac{1}{2} \frac{V_\mathrm{DC}^2 G_e}{1/n_1+1/n_2},
\label{JQ}
\end{equation}
with $(1/n_1+1/n_2)/G_e$ the two terminal resistance across the QPCs and micron-sized ohmic contact. The factor one half with respect to the standard two terminal expression for the total Joule power $V_\mathrm{DC}^2/R$ results from the equal power dissipated into the cold electrodes at temperature $T_0$. Note that there are no such complications as Peltier or thermopower effects for perfectly transmitted channels (see e.g. [8,~22] and the supplementary reference \cite{Molenkamp1990}).\\

\subsection{Noise measurement setup}
A simplified circuit description of the noise measurement apparatus is displayed in supplementary Fig.~S1A, and the practical implementation is shown in supplementary Fig.~S1B.\\

\begin{figure*}[!htbp]
\renewcommand{\thefigure}{\textbf{S\arabic{figure}}}
\renewcommand{\figurename}{\textbf{Figure}}
\centering\includegraphics[width=0.8\textwidth]{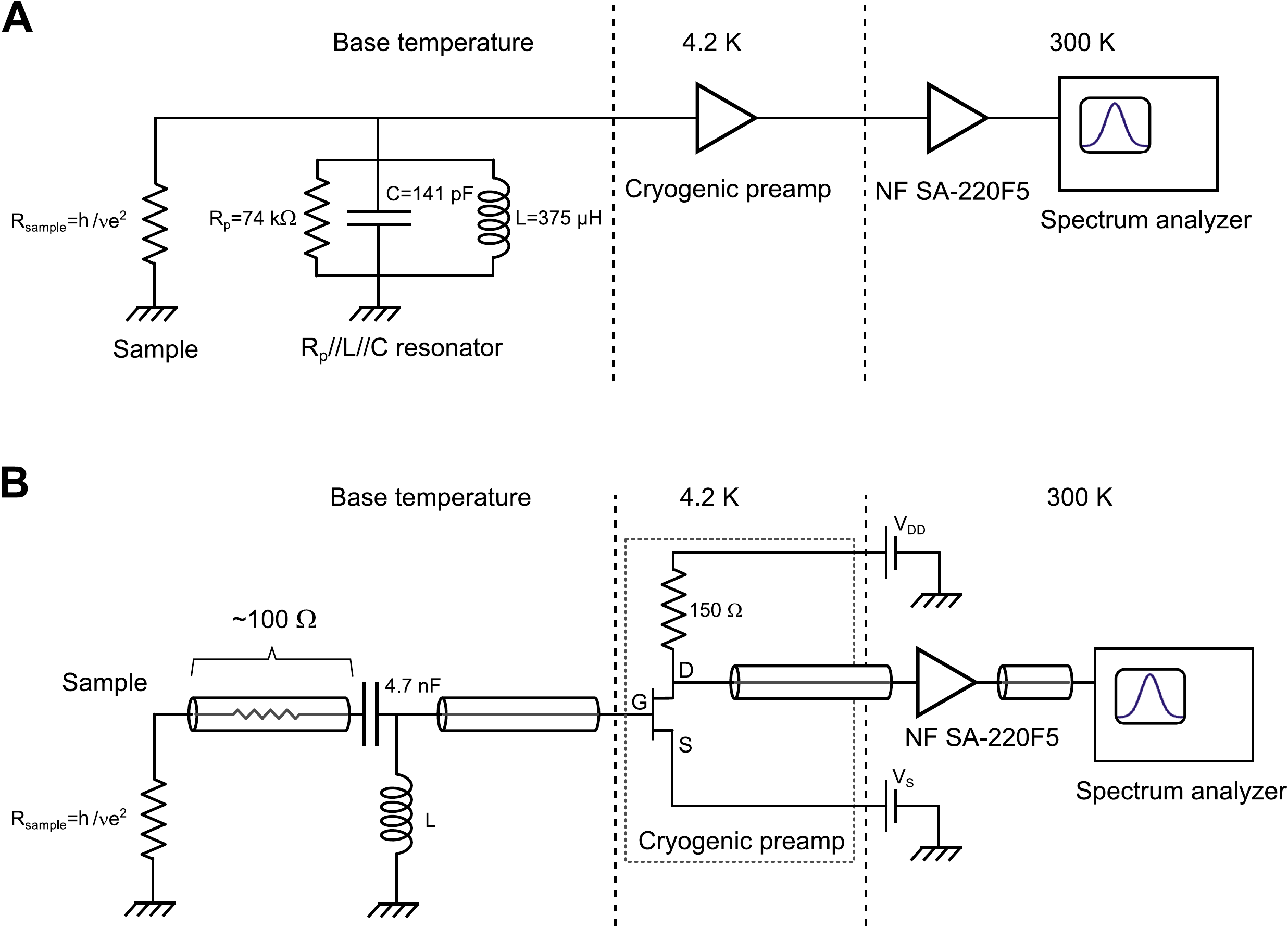}
\caption{
Noise measurement setup. \textbf{(A)} Simplified schematic description: the sample is connected in parallel to a $R_p//L//C$ resonator; the signal is then amplified by an home-made cryogenic preamplifier followed by a room temperature amplifier. The listed $R_p,$ $L,$ $C$ values of the resonator correspond to the measurement chain 2, behind QPC$_2$, which was used for the data shown in the article. \textbf{(B)} Practical implementation: The resonator inductance $L$ is made of a superconducting coil thermally anchored to the mixing chamber. The resonator capacitance $C$ is mostly given by the distributed capacitance along the coaxial cables connecting the sample to the cryogenic preamplifier. The parallel shunt resistor $R_p$ models the dissipation mostly due to the $\sim 100~\Omega$ resistance of the coaxial wire between sample and superconducting coil. Note the presence of a DC block $4.7~$nF capacitance.
}
\label{figSI-schematic}
\end{figure*}

{\noindent\textbf{Amplification chain.}} The measurements are performed with an ultra-low noise preamplifier based on a home-grown high electron mobility transistor [24] thermalized to the $4.2~$K stage of the dilution refrigerator. A simplified schematic of the cryogenic preamplifier is shown in supplementary Fig.~S1B. It is tuned with the DC voltages $V_{DD}$ and $V_S$, while the gate port of the HEMT is DC-grounded through the superconducting coil (see supplementary Fig.~S1B). In the investigated configurations, the cryogenic preamplifier input noise was $\sim 0.2~\mathrm{nV}/\sqrt{\mathrm{Hz}}$ (see supplementary Fig.~S3D for the measured input noise spectral density of the full amplification chain) and its voltage gain was $\sim 5$.

After a second stage of amplification at room temperature (amplifier NF SA-220F5), the signal is digitized at $10~$MS/s (NI PXI-5922) to compute the noise spectral density.

The noise spectral density is then integrated around the resonance over a finite bandwidth, which is tuned separately for each filling factor in order to optimize the signal-to-noise ratio.\\

{\noindent\textbf{Resonant circuit.}} A resonant $L//C$ tank (see supplementary Fig.~S1) is used to shift the working frequencies in the MHz range (near 0.7~MHz), where the cryogenic preamplifier shows the best performance. Another motivation is that many extrinsic noise sources become negligible in this frequency range, such as low-frequency charges trapping in the substrate or vibration noise.

The inductance $L$ is realized by a superconducting coil thermalized to the mixing chamber of the dilution refrigerator. The parallel capacitance $C$ is mostly due to the distributed capacitance along the coaxial lines connecting the sample to the cryogenic preamplifier. To model the dissipation between the output of the sample and the input of the amplifier we introduced a parallel resistance $R_p$ to the tank. Here the dissipation mostly results from the resistive coaxial wire between the sample and the superconducting coil (see supplementary Fig.~S1B). The precise check of this simplified model for the resonant circuit and the determination of the resonant circuit parameters is detailed below, in the corresponding subsection of "Calibrations" ($R_p\simeq 94.5 \pm 2~\mathrm{ k} \Omega$ for amplification chain 1, behind QPC$_1$ and $R_p\simeq 73.7 \pm 1.5~\mathrm{ k} \Omega$ for amplification chain 2, behind QPC$_2$). The $4.7~$nF capacitance between sample and inductor (see supplementary Fig.~S1B) blocks the DC current along the measurement line and isolates the DC voltage at the preamplifier gate port from the sample bias voltage.\\

\subsection{Calibrations}

{\noindent\textbf{Amplification chain.}}
The overall amplification chain (including both cryogenic and room temperature preamplifiers) is calibrated from the temperature dependent Johnson-Nyquist noise down to $50~$mK following e.g. \cite{DiCarlo2006}, see supplementary Fig.~S2. The calibration was repeated for each filling factor $\nu\in\{3,4\}$, each corresponding to a different sample impedance $h/\nu e^2$ at the input of the preamplifier and to a different choice of bandwidth $\delta$f for the integration of the measured noise spectral density. The integrated noise measured at the output of the full amplification chain (at the output of the room temperature amplifier) is plotted in supplementary Fig.~S2 as a function of temperature for the two filling factors $\nu=3$ and $\nu=4$.

\begin{figure}[!htbp]
\renewcommand{\thefigure}{\textbf{S\arabic{figure}}}
\renewcommand{\figurename}{\textbf{Figure}}
\centering\includegraphics[width=0.9\columnwidth]{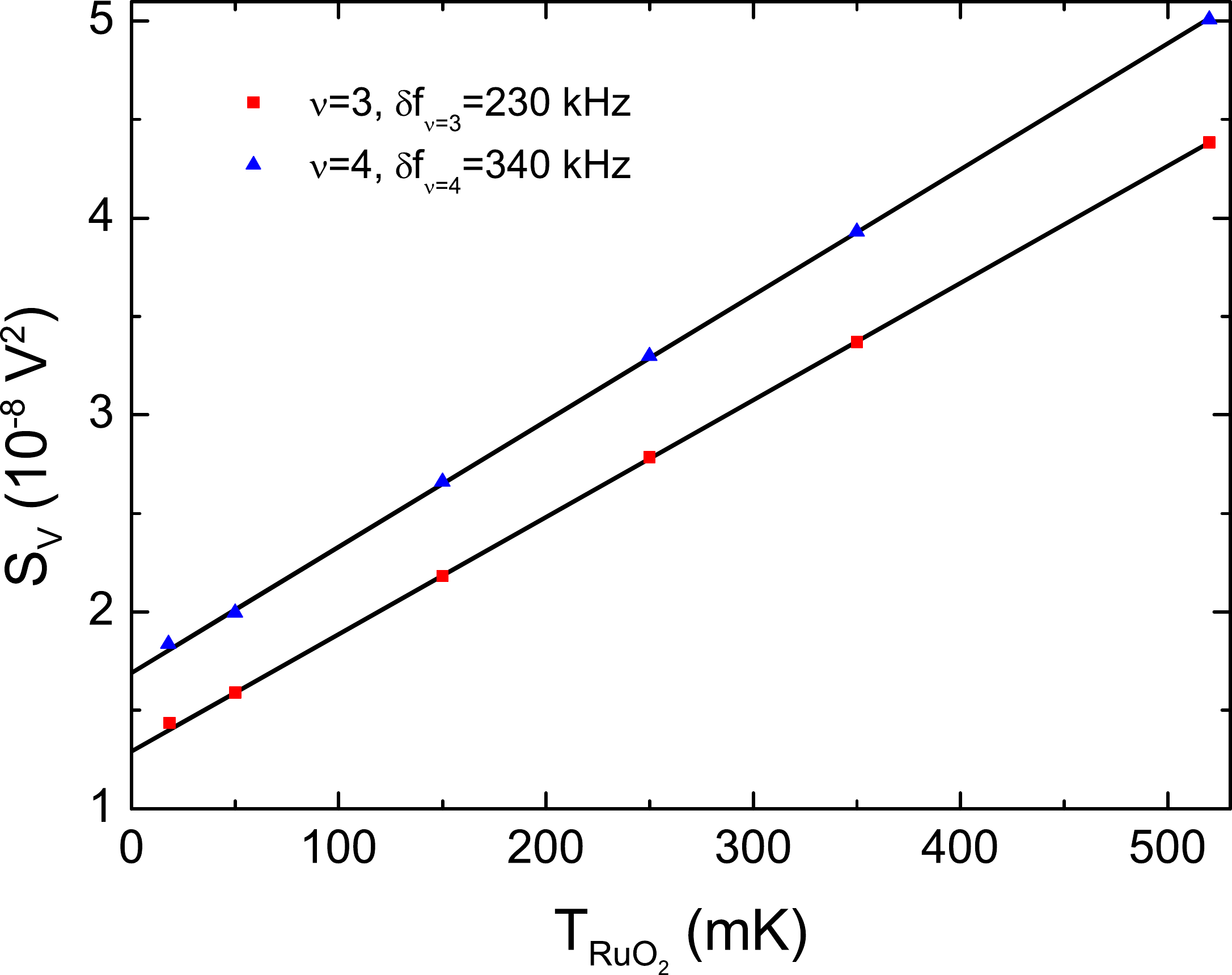}
\caption{
Amplification chain calibration. Symbols correspond to the noise measured at the output of the full amplification chain after integration over the measurement bandwidth $\delta$f, for the two investigated filling factors $\nu=3,4$. The data is plotted versus the temperature readings $T_{\mathrm{RuO}_2}$ of a RuO$_2$ sensor thermalized to the mixing chamber.
Continuous lines are linear fits of the data at $T_{\mathrm{RuO}_2}\geq 50~$mK. Essentially, the gain and the system noise offset can be inferred from, respectively, the slope and the intersect at $T=0$. Note that the larger offset at filling factor $\nu=4$ mostly results from the choice of a broader integration bandwidth $\delta$f.
}
\label{figSI-calibint}
\end{figure}

Note that the calibrated overall gain matches the less accurate procedure using the known gain of the room temperature amplifiers ($\sim 200$) and the cryogenic preamplifiers gain ($\sim 5$) obtained by the in-situ measurements of its DC transconductance and output impedance [24]. Note also that the RuO$_2$ temperature sensor is located outside the magnet bore. It is therefore subjected to a much smaller magnetic stray field than the sample.

\begin{figure*}[!htbp]
\renewcommand{\thefigure}{\textbf{S\arabic{figure}}}
\renewcommand{\figurename}{\textbf{Figure}}
\centering\includegraphics[width=0.8\textwidth]{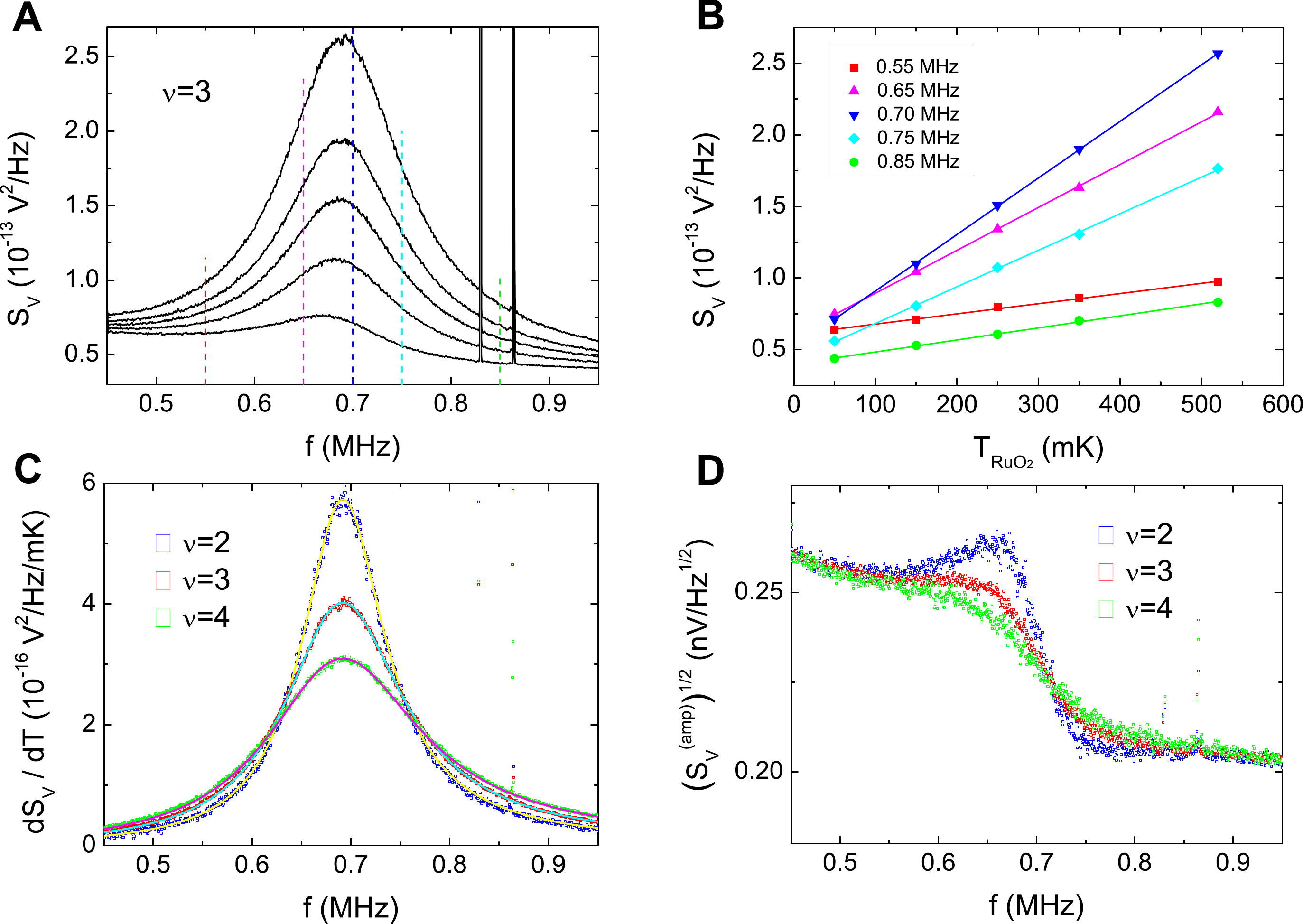}
\caption{
Characterization of the resonant circuit.
\textbf{(A)} Equilibrium voltage noise spectrum measured at the output of the amplification chain 2, for filling factor $\nu=3$. Each trace corresponds to a different temperature $T_{\mathrm{RuO}_2}$: from bottom to top, 50, 150, 250, 350 and 520 mK.
\textbf{(B)} Linearity of the noise as a function of temperature. The symbols represent the extracted values of the above noise spectra at five different frequencies, indicated by the dashed lines in the previous panel. The continuous lines are linear fits of these data.
\textbf{(C)} Slope of the temperature dependence of the noise spectrum, at filling factors $\nu=2,3,4$. The symbols are the slopes obtained by applying the linear fit procedure shown in panel \textbf{B} to every frequency of the spectrum. The continuous lines are fits using supplementary Eq.~\ref{EqCalib} and the model shown in supplementary Fig.~S1A.
\textbf{(D)} Voltage noise added by the amplification chain, referenced to the input. The symbols are extracted from the $T=0$ intercept of the linear fits, at filling factors $\nu=2,3,4$.
}
\label{figSI-CalibTank}
\end{figure*}

Subsequently, we checked during the measurements the stability of the overall amplification chain gain and noise offset. The gain stability is monitored with a fixed AC signal slightly outside the frequency range used for the noise measurement (seen as high and narrow spikes in supplementary Fig.~S3A). The overall system noise is monitored from the noise signal at much higher frequencies, where the $R_p//L//C$ tank is essentially a short circuit. In practice, we found that the amplification chains remained stable at the experimental accuracy.\\

{\noindent\textbf{Base sample temperature.}}
We take advantage of the overall amplification chain calibration presented above to precisely extract the base temperature on the sample $T_0$ at the lowest temperatures. To do so, we measure the Johnson-Nyquist noise of the sample below 50~mK and use the known amplification gain to infer the temperature.
We find that our RuO$_2$ thermometer, directly attached to the mixing chamber of the dilution refrigerator, deviates from the noise thermometer at the lowest temperatures by a relatively small amount, smaller than 5~mK (see supplementary Fig.~S2). There are several possible explanations for this relatively small discrepancy. One is the calibration of the RuO$_2$ thermometer, which is not accurate below $40~$mK. Another is that the sample itself is slightly warmer than the mixing chamber, for instance due to spurious power sources. Note however that the temperature extracted from the noise measurement remains the pertinent electronic temperature. Note also that the base temperature $T_0$ was extracted separately at each filling factor $\nu$. \\

{\noindent\textbf{Resonant circuit.}}
Here we present a full characterization of the measurement circuit, upstream of the cryogenic preamplifier. We demonstrate that this resonant circuit is accurately described by the $R_p//L//C$ resonator depicted in supplementary Fig.~S1A, and determine the value of these parameters. For this purpose, the full frequency dependent voltage noise was measured at different temperatures and for different values of the sample resistance, at $\nu=2$, 3 and 4.

The symbols in supplementary Fig.~S3A represent the equilibrium voltage noise spectrum $S_V(f)$ measured at the output of the amplification chain for several temperatures and at filling factor $\nu=3$. At equilibrium, the measured voltage noise is given by the Johnson-Nyquist noise of the resonant circuit in parallel to the sample, and by the noise added by the amplifiers $S_V^\mathrm{(amp)}$ (mainly due to the cryogenic preamplifier):
\begin{equation}
S_V(f) = G^2 \left( 4 k_B T \mathrm{Re}\left[Z(f)\right] + S_V^\mathrm{(amp)}(f) \right),
\label{EqCalib}
\end{equation}
where $G$ is the total gain of the amplification chain and $Z(f)$ is the measurement impedance $(h/\nu e^2)//R_p//L//C$ in the model depicted in supplementary Fig.~S1A.

In order to accurately characterize the resonant circuit, one needs to cancel out the amplifier contribution $S_V^\mathrm{(amp)}$. This can be done by changing the temperature, since only the resonant circuit noise contribution depends on temperature.

As illustrated in supplementary Fig.~S3B for five different frequencies, we find as expected that the measured output noise spectrum is a linear function of the temperature. We perform a linear fit of the temperature dependence at each frequency of the spectrum. The obtained slopes are then plotted as a function of the frequency in supplementary Fig.~S3C, together with the slopes similarly extracted at filling factors $\nu=2$ and $4$. A simultaneous fit of those three datasets (shown as continuous lines in supplementary Fig.~S3C) unambiguously yields the resonant circuit parameters $R_p$, $L$, $C$ shown in supplementary Fig.~S1A, as well as the overall gain $G$. The high quality of the fits over a broad frequency range validates unambiguously the simple $R_p//L//C$ model of the resonant circuit.

Interestingly, the amplifier noise $S_V^\mathrm{(amp)}(f)$ can be extracted from the $T=0$ intercepts of the linear fits. The results are shown in supplementary Fig.~S3D for the three filling factors $\nu=2,3$ and 4. The noise at resonance depends on the filling factor because of the small contribution of the preamplifier current noise.

{\subsection{Expression of the noise spectral density with an increased $\Tohm$.}}
We now derive the relation used to extract $\Tohm$ from the measured noise spectral density (article Eq.~3). We follow a standard approach using the Landauer-B\"{u}ttiker scattering description (see e.g. [27] and the supplementary references \cite{Buttiker1992,Blanter2000,Texier2000}). An identical result for the excess current noise was obtained in [28].

The current fluctuations $\delta I_i$ \textit{emitted} by the floating micron-sized ohmic contact across one of the $n$ open electronic channels, labeled $i$, is the sum of two distinct contributions:
\begin{eqnarray}
\delta I_i=\delta I_i^{\Tohm}+\delta V_\Omega G_e.
\end{eqnarray}
The first contribution $\delta I_i^{\Tohm}$ corresponds to the Fermi distribution of the current carrying outgoing states population, whose associated noise spectral density reads $<(\delta I_i^{\Tohm})^2>=2k_B\Tohm G_e$.
The second contribution $\delta V_\Omega G_e$ corresponds to the voltage fluctuations of the micron-sized ohmic contact. Since in the present sample the capacitance to ground of the floating micron-sized ohmic contact is very small ($C\simeq 2.3$~fF from numerical computation and dynamical Coulomb blockade measurements on the same sample, see [26]; this capacitance corresponds to a parallel impedance of approximately 100~M$\Omega$ in the investigated frequency range), the overall incoming and outgoing currents must match:
\begin{eqnarray}
n\delta V_\Omega G_e+\sum_{i=1}^{n} \delta I_i^{\Tohm}=\sum_{i=1}^{n} \delta I_i^{T_0},
\end{eqnarray}
where $\delta I_i^{T_0}$ is the incoming current fluctuation at the micron-sized ohmic contact, which was transmitted across the open electronic channel $i$ and emitted from a large cold electrode at $T_0$ ( $<(\delta I_i^{T_0})^2>=2k_BT_0 G_e$).

Summing up the contributions of the $n_1$ ($n_2$) open electronic channels across QPC$_1$ (QPC$_2$) gives the spectral density of the emitted current noise toward the measurement electrode behind QPC$_1$ (QPC$_2$). The overall current noise spectral density is found independent of the measurement electrode and simply given by:
\begin{eqnarray}
S_I = \frac{2k_B (\Tohm-T_0) G_e}{1/n_1 + 1/n_2}+4k_B T_0 \nu G_e.
\label{eq-FullNoise}
\end{eqnarray}\\

\subsection{Test of the noise measurement setup: comparison of excess noise from the two amplification chains.}

This comparison demonstrates the relative precision of our calibration procedure, and validates a prediction of supplementary Eq.~\ref{eq-FullNoise}, namely that the noise spectral density is identical at the two measurement electrodes, independently of the number of open channels in QPC$_1$ and QPC$_2$.

\begin{figure}[!htbp]
\renewcommand{\thefigure}{\textbf{S\arabic{figure}}}
\renewcommand{\figurename}{\textbf{Figure}}
\centering\includegraphics[width=0.9\columnwidth]{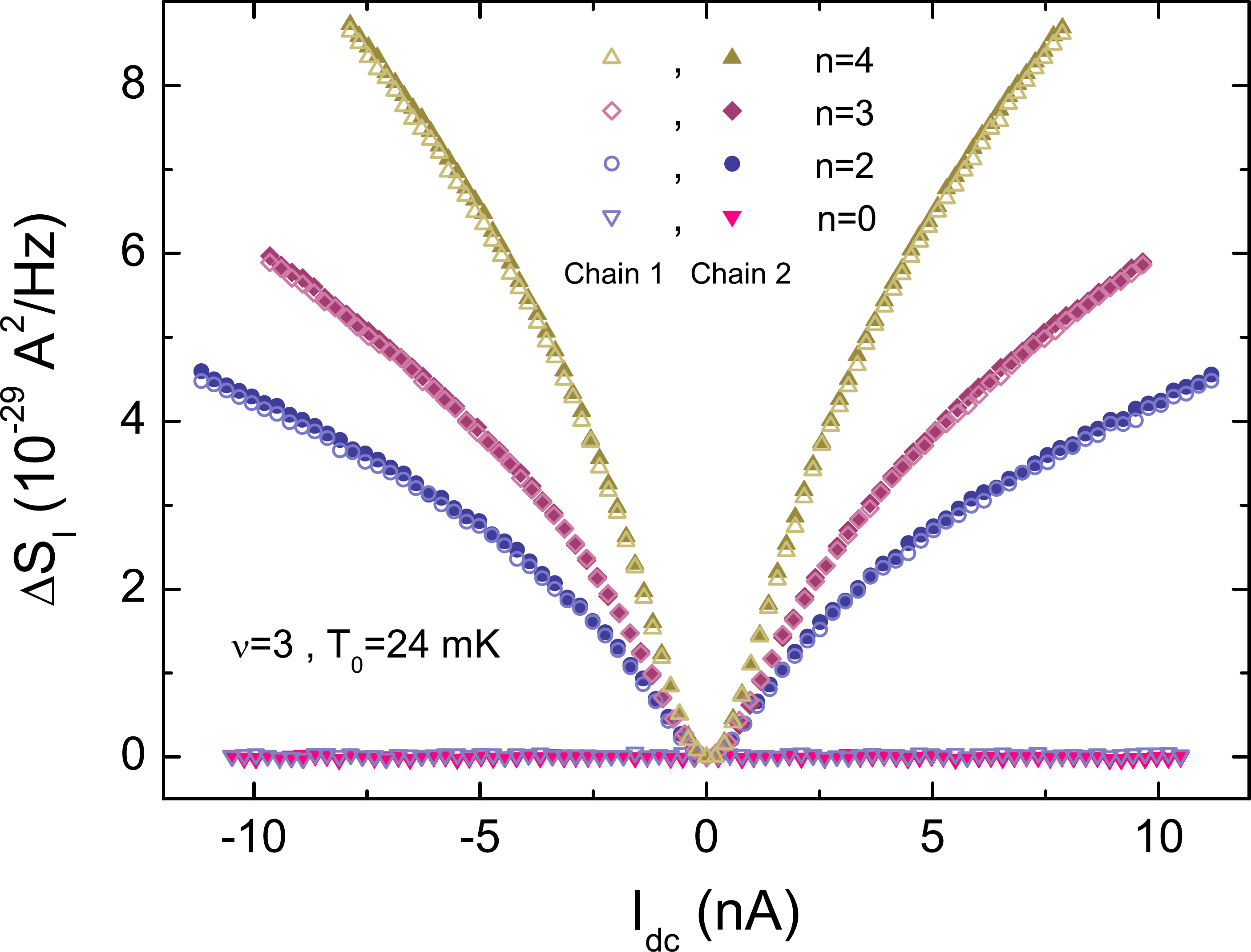}
\caption{
Comparison of the noise measurements from both amplification chains. Measured noise spectral density at $\nu=3$ and $T_0=24~$mK versus injected DC current. The symbols correspond (from bottom to top) to $n=0$, 2, 3 and 4 open channels. Open symbols: measurement with amplification chain~1. Filled symbols: measurement with amplification chain~2.
}
\label{figSI-hemt1and2}
\end{figure}

Two separate amplification chains are used to measure the noise, each connected to one of the two measurement electrodes (see article Fig.~1B). Chain 1 is connected to the measurement electrode located behind $\QPCun$, and chain 2 to the measurement electrode behind $\QPCdeux$. According to theory, the excess noise is the same at either output, and is given by article Eq.~3. The data measured with the two amplification chains agree within less than $2\%$, as illustrated in supplementary Fig.~S4.\\

\subsection{The quasi-equilibrium hypothesis for electrons in the heated-up metal plate.}

We here estimate the dwell time of an electron in the micron-size ohmic contact and compare it to the typical electron-electron interaction time. As a result of the very large electronic density of states in the metallic micron-size ohmic contact, the dwell time $t_\mathrm{dwell}$ is found in the range of $10~\mu$s, very much larger than the typical electron-electron interaction time, which is in the 10~ns range at low temperatures (see e.g. [30]) for the related measurement of $\tau_\phi$ in gold). This firmly establishes the quasi-equilibrium hypothesis that the electrons energy distribution in the micron-size ohmic contact is a hot Fermi distribution characterized by a temperature $\Tohm$.

The electronic dwell time $t_\mathrm{dwell}$ inside the micron-size ohmic contact is evaluated from the classic expression $t_\mathrm{dwell}= \nu_F \Omega h / n$ (see e.g. \cite{Brower1997}), with $\Omega$ the volume of the micron-size ohmic contact and $\nu_F$ the electronic density of states per unit volume and energy. Injecting the volume of the micron-size ohmic contact $\Omega \approx 2~\mu\mathrm{m}^3$ and a typical density of states for metals $\nu_F\approx 10^{47}~\mathrm{J}^{-1}\mathrm{m}^{-3}$ (in gold $\nu_F \simeq 1.14 \times 10^{47}~\mathrm{J}^{-1}\mathrm{m}^{-3}$), we find $t_\mathrm{dwell}\simeq \frac{60}{n}~\mu$s.\\

\subsection{Electron-phonon heat flow.}

We now discuss the assumptions made in the model dependent analysis of the $n=4$ reference data. In this analysis, we assume the following standard expression for the electron-phonon heat flow: $\Jeph(\Tohm,T_0)=\Sigma\Omega(\Tohm^5-T_0^5)$. Note that our main result summarized article Eq.~4 does not rely on this model (or any other), since we extracted separately the electronic heat flow by changing the number $n$ of open electronic channels.\\

{\noindent\textbf{Cold phonon hypothesis.}}
The above expression assumes that the temperature of the phonons coupled to the electrons in the micron-sized ohmic contact is the same as the measured electronic temperature $T_0$.

Although this is a standard hypothesis for such low levels of injected power in similar devices at low temperatures (see e.g. [31] and references therein), we confirmed its validity in the present experimental setup by extracting the heat flow at different values of the temperature $T_0$:
As pointed out in the article, we obtain the same value for the electron-phonon prefactor $\Sigma\Omega$ when fitting the reference $n=4$ data measured either at $T_0=24~$mK or $T_0=40~$mK. This shows that the phonons are not significantly heated up. Indeed an actual phonon temperature significantly higher than $T_0$ would lead to different values of $\Sigma\Omega$ as the base temperature is changed. Note that this is a commonly used criterion to demonstrate the presence of hot phonons \cite{Rajauria2007}.\\

{\noindent\textbf{Temperature exponent.}}
The above standard expression for the electron-phonon heat flow was verified in most experiments involving similar metal devices at low temperatures (see e.g. [31] and references therein). Nevertheless, deviations to the $T^5$ power law were also observed and predicted, for instance in highly disordered metals (see e.g. \cite{Sergeev2000} and references therein). These deviations usually take the form of a different temperature power law $T^p$, where the exponent $p$ varies between 3 and 6.

In the article we focus on the low injected power regime $J_Q\lesssim 20$~fW where electronic heat flow is dominant, and we assume an exponent $p=5$.
We tested the validity of this hypothesis by fitting the same data using the temperature exponent $p$ as a free parameter. In order to focus on the electron-phonon coupling, which becomes dominant at high injected power, the fit is performed up to the largest injected power $J_Q\sim 90$~fW. We found that the best fit is obtained for the exponent $p=4.8$ (both for the $\nu=3$ and $\nu=4$ reference data), with a fitted electronic heat flow $\alpha_4$ found $1\%$ higher at $\nu=4$ ($11\%$ lower at $\nu=3$) than the predicted quantum limit. This confirms that the electron-phonon heat flow follows closely the above standard expression.\\

\subsection{Experimental uncertainty on the quantum limit of heat flow.}

\begin{table}[ht]
\centering
\renewcommand{\thetable}{\textbf{S\arabic{table}}}
\renewcommand{\tablename}{\textbf{Table}}
 \renewcommand{\arraystretch}{1.4}
\begin{tabular}{|c||c|c|c||c|c|c|c|c|}
\hline
 & \multicolumn{3}{c||}{$\nu=3$} & \multicolumn{5}{c|}{$\nu=4$}\\
\hline
 $n$ & 2 & 3  & 4 & 2 & 3  & 4 & 5 & 6   \\
\hline\hline	
$\alpha_{n-4}'$ & -2.00 & -0.93 &   & -2.33 & -1.40 &   & 1.06 & 1.83 \\
%\hline	
%$\frac{\alpha_{n-4}'}{n-4}$ & 1.00 & 0.93 &   & 1.16 & 1.40 &   & 1.06 & 0.92 \\
\hline\hline	
$\alpha_{n}$& 1.78 & 2.85 & 3.78  & 1.84 & 2.77 & 4.17  & 5.23 & 6.00 \\
%\hline	
%$\frac{\alpha_{n}}{n}$ & 0.89 & 0.95 & 0.95  & 0.92 & 0.92 &  1.04 & 1.05 & 1.00 \\
\hline
\end{tabular}
\caption{Extracted values $\alpha_{n-4}'$ (model-free approach) and $\alpha_{n}$ (model-dependent approach).} \label{table:alphan}
\end{table}

Here we provide further details on the uncertainties estimation regarding the experimentally determined quantum limit of heat flow across one electronic channel (the uncertainties displayed article Eq.~4 and in the paragraph below article Eq.~4).

\begin{figure}[!tbh]
\renewcommand{\thefigure}{\textbf{S\arabic{figure}}}
\renewcommand{\figurename}{\textbf{Figure}}
\centering\includegraphics[width=\columnwidth]{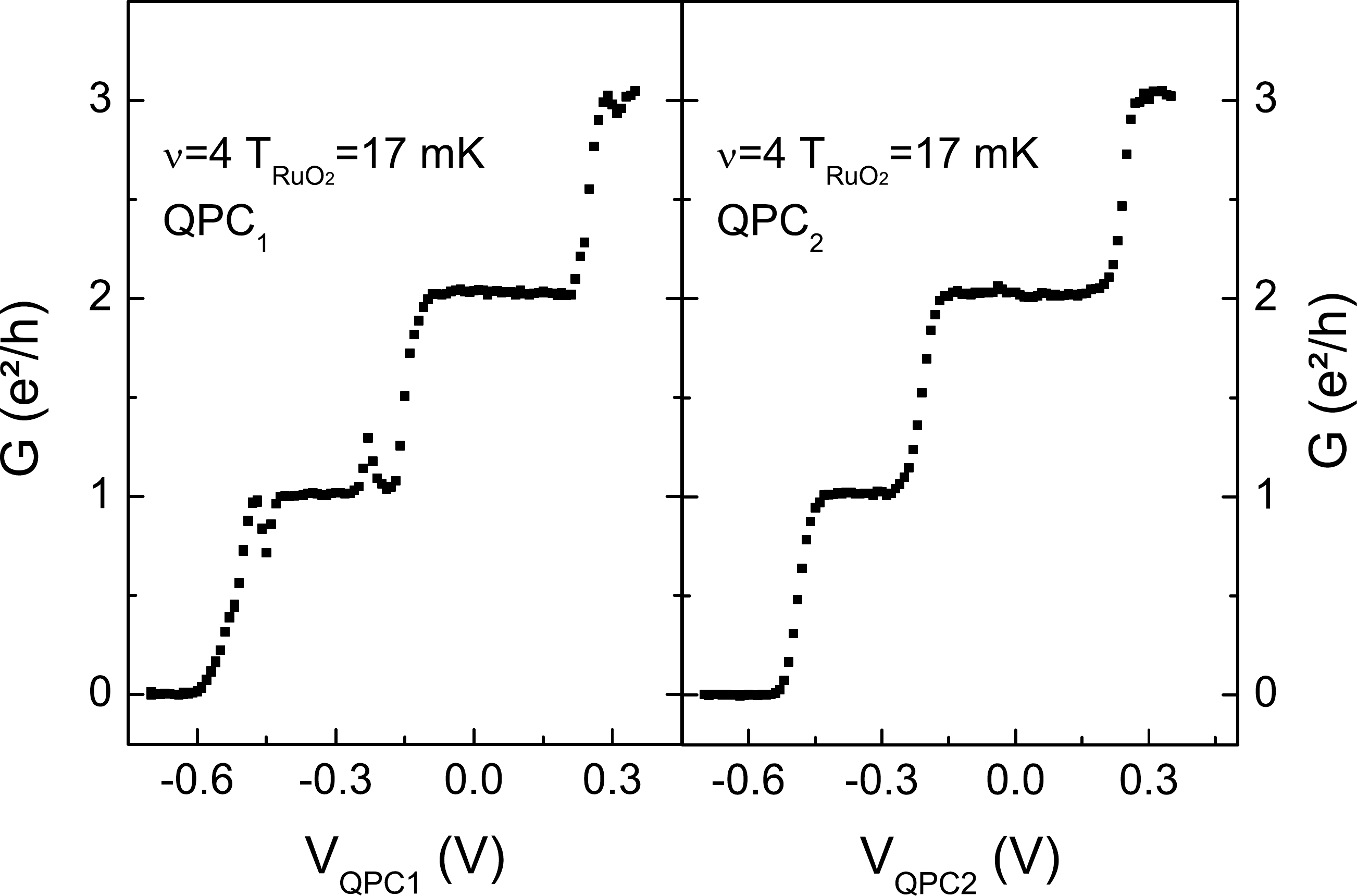}
\caption{
Typical conductance traces measured at $\nu=4$ for $\QPCun$ (left panel) and $\QPCdeux$ (right panel), plotted as a function of the corresponding applied gate voltage.
}
\label{figSI-QPCs}
\end{figure}

\begin{figure}[!htbp]
\renewcommand{\thefigure}{\textbf{S\arabic{figure}}}
\renewcommand{\figurename}{\textbf{Figure}}
\centering\includegraphics[width=0.8\columnwidth]{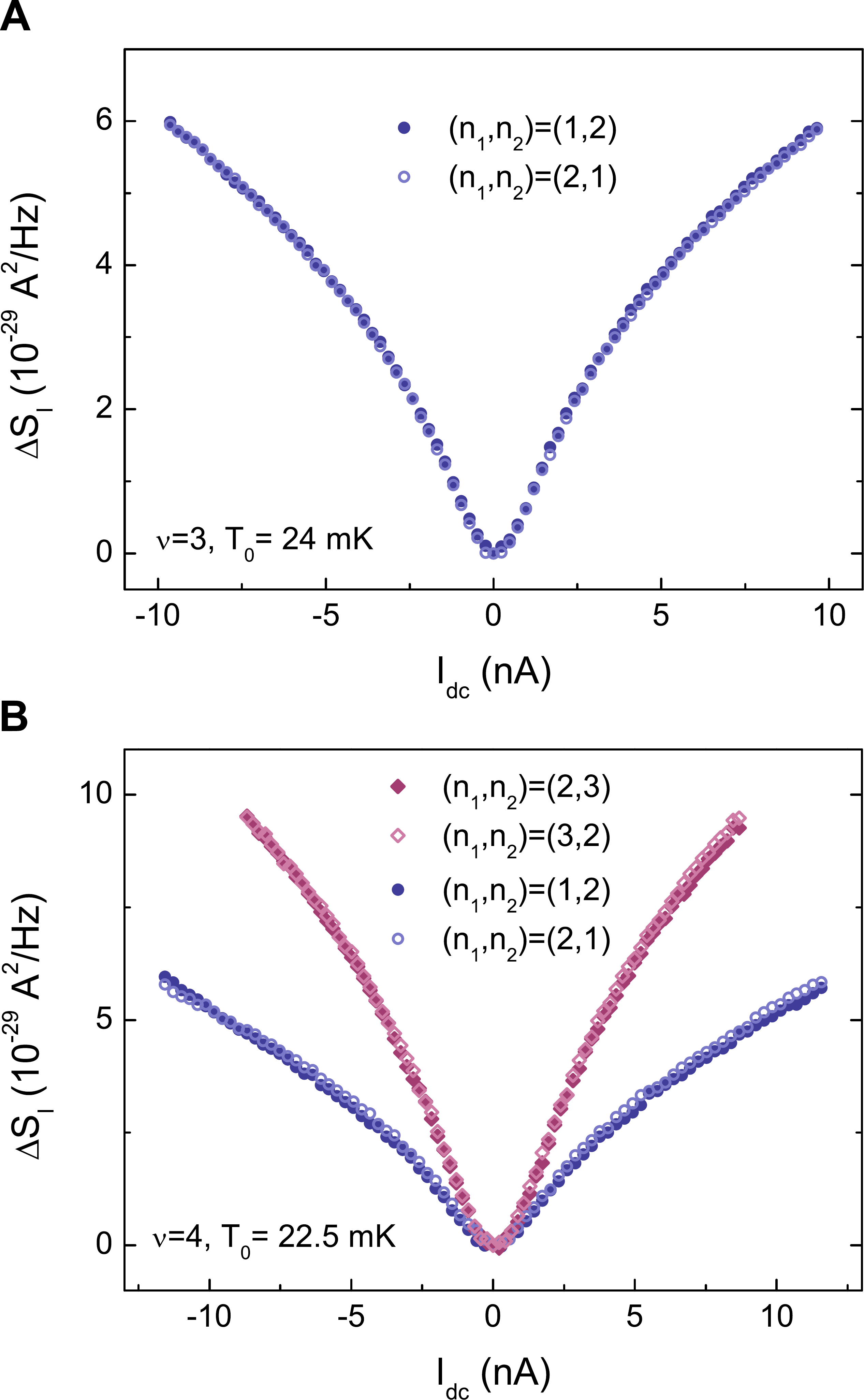}
\caption{
Comparison of measured noise for symmetric sample configurations. \textbf{(A)} Measured noise spectral density at $\nu=3$ and $T_0=24~$mK versus injected DC current, for the $(n_1=1,n_2=2)$ (filled symbols) and $(n_1=2,n_2=1)$ (open symbols) configurations. \textbf{(B)} Measured noise spectral density at $\nu=4$ and $T_0=22.5~$mK versus injected DC current, for the $(n_1=1,n_2=2)$ (filled circles), $(n_1=2,n_2=1)$ (open circles), $(n_1=2,n_2=3)$ (filled diamonds) and $(n_1=3,n_2=2)$ (open diamonds) configurations.
}
\label{figSI-symmetricconfigs}
\end{figure}

\begin{figure}[!htb]
\renewcommand{\thefigure}{\textbf{S\arabic{figure}}}
\renewcommand{\figurename}{\textbf{Figure}}
\centering\includegraphics[width=\columnwidth]{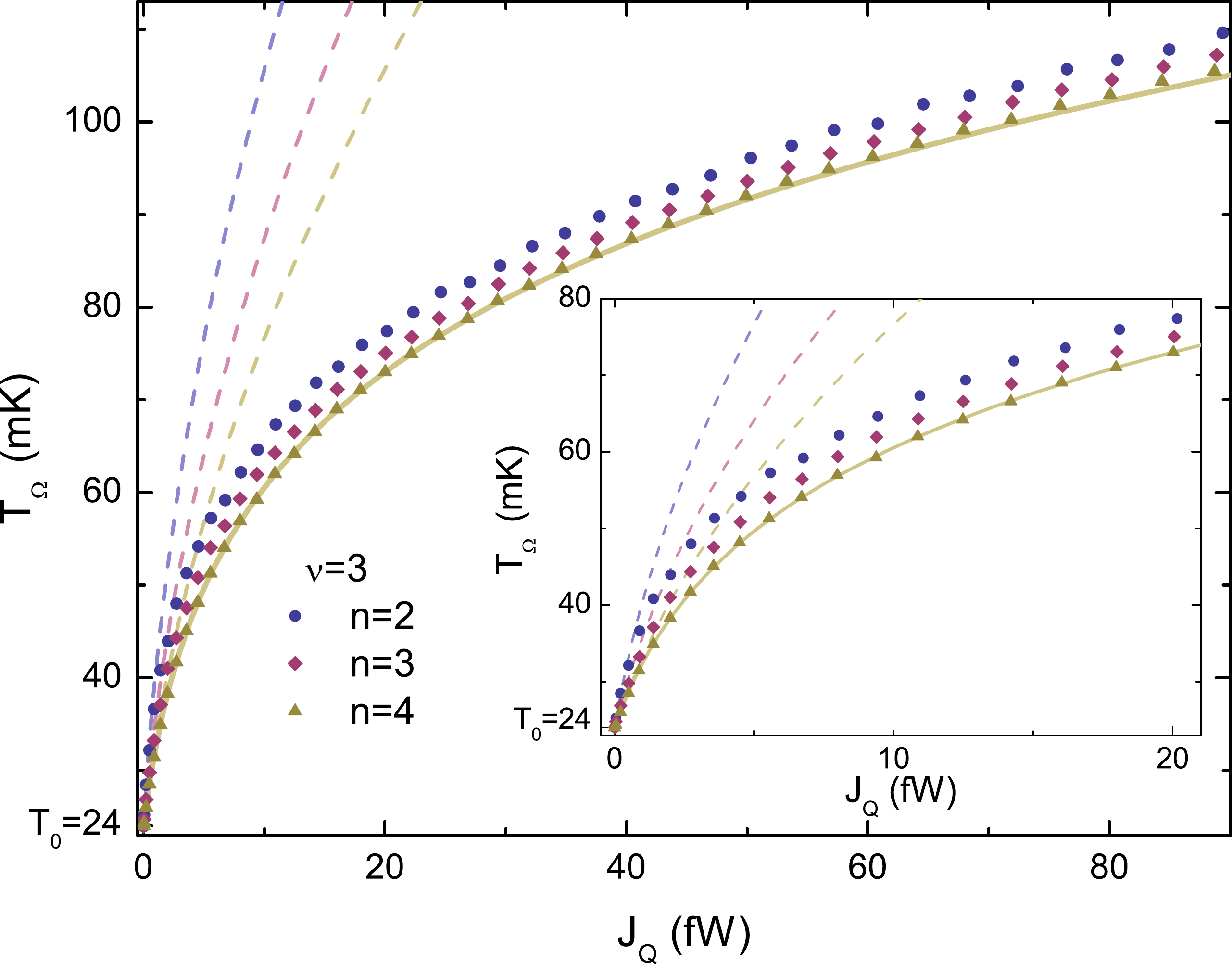}
\caption{
Noise measurements on the full span of injected power $J_Q$, at $\nu=3$ and $T_0=24~$mK. The electronic temperature $\Tohm$ in the micron-sized ohmic contact is plotted as a function of the injected power. The different set of symbols correspond (from top to bottom) to $n=2$, 3 and 4 open channels. The size of the symbols indicates the uncertainty. The continuous line is a fit of the reference data for $n=4$ open channels including the heat transfer toward phonons. Note that the fit was performed in the low injected power range $J_Q\lesssim 20$~fW where the electronic heat flow is dominant, and evaluated with the same parameter up to the maximum injected power $J_Q\simeq90~$fW. The dashed lines correspond to the expected contribution of the electronic heat flow $n\Je$ in absence of electron-phonon cooling. Inset: zoom on the data for the restricted span of $\Jin$ shown in the main text, emphasizing the fact that on this span, electronic heat flow and electron-phonon cooling have similar contributions.
}
\label{figSI-datanu3}
\end{figure}

\begin{figure}[!htbp]
\renewcommand{\thefigure}{\textbf{S\arabic{figure}}}
\renewcommand{\figurename}{\textbf{Figure}}
\centering\includegraphics[width=\columnwidth]{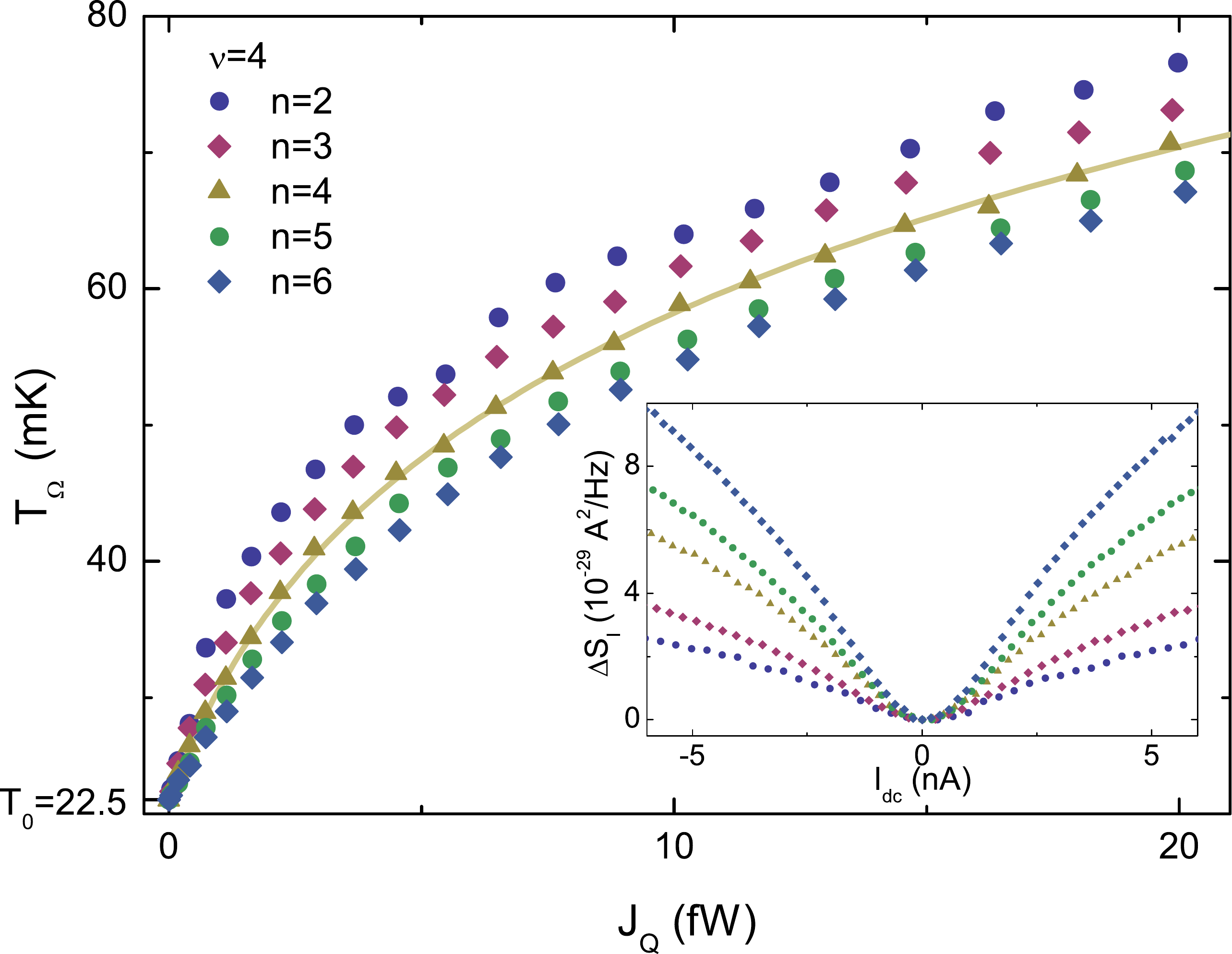}
\caption{
Noise measurements at $\nu=4$ and $T_0=22.5~$mK. Variation of the electronic temperature in the micron-sized ohmic contact $\Tohm$ as a function of the injected power, at $\nu=4$ and $T_0=22.5~$mK. The symbols correspond (from top to bottom) to $n=2$, 3, 4, 5 and 6 open channels. The size of the symbols indicates the uncertainty. The continuous line is a fit of the reference data for $n=4$ open channels, including the heat transfer toward phonons. Inset: raw noise data versus injected DC current.
}
\label{figSI-datanu4}
\end{figure}

\begin{figure*}[!htbp]
\renewcommand{\thefigure}{\textbf{S\arabic{figure}}}
\renewcommand{\figurename}{\textbf{Figure}}
\centering\includegraphics[width=0.9\textwidth]{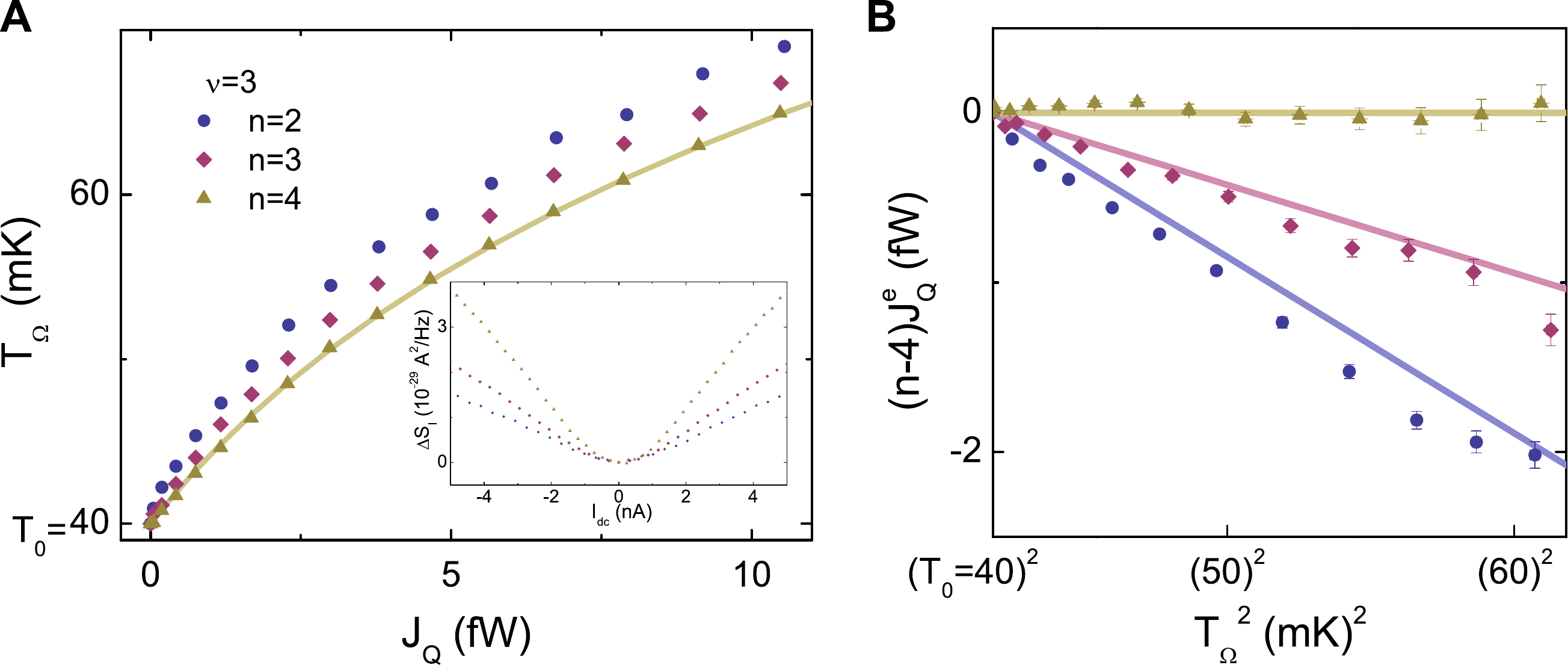}
\caption{
Noise measurements and corresponding electronic heat current at $\nu=3$ and $T_0=40$~mK. \textbf{(A)} Variation of the electronic temperature in the micron-sized ohmic contact $\Tohm$ as a function of the injected power. The symbols correspond (from top to bottom) to $n=2$, 3, and 4 open channels. The continuous line is a fit of the reference data for $n=4$ open channels. Inset: raw noise data versus injected DC current. \textbf{(B)} Symbols display the heat current across $|n-4|$ electronic channels, with a positive (negative) sign for $n>4$ ($n<4$), versus squared temperature, at $T_0=40~$mK and filling factor $\nu=3$, for (from bottom to top) $n=2$, 3 and 4 open channels. The continuous lines are the theoretical predictions for the quantum limited heat flow.
}
\label{figSI-data40mK}
\end{figure*}

We opted for a very straightforward approach: taking advantage of the fact that we have several different measurements of this quantum limit, corresponding to different experimental configurations including the different investigated filling factors $\nu=3$ and $\nu=4$, we calculated for this ensemble of measurements not only the mean value but also but also the standard error on this mean value, which is the displayed uncertainty. The limitation of such an approach is that it ignores systematic errors, and in particular the error on the gain of the amplification chain. Note however that we found that the two amplification chains, calibrated separately, agree within less than $2\%$ (see supplementary Fig.~S4).

In the model-free approach, this analysis is performed on 6 different data points $\{\alpha_{n-4}'/(n-4)\}$ (see supplementary Table~S1, we recall that $\alpha_{n-4}'\equiv(n-4)\Je(\Tohm,T_0)/(\Tohm^2-T_0^2)$), each weighted by the number of probed electronic channels $|n-4|$. This gives:
\begin{eqnarray}
\frac{\Je(\Tohm,T_0)}{\Tohm^2-T_0^2} = (1.06 \pm 0.07)\times\frac{\pi^2 k_B^2}{6h}.
\end{eqnarray}

In the model-dependent approach, this analysis is performed on 8 different data points $\{\alpha_{n}/n\}$ (see supplementary Table~S1, we recall that $\alpha_n\equiv \alpha_{n-4}'+\alpha_4$ with $\alpha_4\equiv 4\Je(\Tohm,T_0)/(\Tohm^2-T_0^2)$), each weighted by the number $n$ of open electronic channels. This gives:
\begin{eqnarray}
\frac{\Je(\Tohm,T_0)}{\Tohm^2-T_0^2} = (0.98 \pm 0.02)\times\frac{\pi^2 k_B^2}{6h}.
\end{eqnarray}

\subsection{Supplementary data}

{\noindent\textbf{QPC tuning.}}
The QPCs are tuned to a set of fully open channels by adjusting the bias voltage applied to the corresponding metal split gates.
We show in supplementary Fig.~S5 typical conductance traces of the two QPCs measured at the filling factor $\nu = 4$. These traces display clear conductance plateaus located at integer multiples of $e^2/h$.\\

{\noindent\textbf{Comparison of symmetric sample configurations.}}
In the article, the data for symmetric sample configurations (such as $(n_1=1,n_2=2)$ and $(n_1=2,n_2=1)$, or $(n_1=2,n_2=3)$ and $(n_1=3,n_2=2)$) are averaged in order to improve the signal to noise ratio. In supplementary Fig.~S6, we show that the increase $\Delta S_I$ in the measured spectral density of the current noise is identical, at our experimental accuracy, for symmetric sample configurations, as expected from article Eq.~3 and supplementary Eq.~\ref{eq-FullNoise}.\\

{\noindent\textbf{Measured $T_\Omega$ on the full span of injected power $J_Q$.}}
The data shown in the article are restricted to the low injected power regime $J_Q\lesssim 20$~fW, where the contribution of the electronic channels to the overall outgoing heat flow is dominant. In the main panel of supplementary Fig.~S7, we display as symbols the same data obtained at $\nu=3$ and $T_0=24~$mK and showed in the main panel of article Fig.~2, but up to the maximum injected power $J_Q\simeq90$~fW. Note that these data correspond to the raw noise measurements shown in supplementary Fig.~S4.
Up to the largest injected powers, electron-phonon heat transfers remain well accounted for by the same coupling parameters obtained by fitting the data at $J_Q < 20$~fW (see main text). The continuous line in supplementary Fig.~S7 is calculated with the fit parameters obtained at low injected power for the $n=4$ reference configuration. The dashed lines correspond to the expected contribution of the electronic heat flow $n\Je$ in absence of electron-phonon cooling.\\

{\noindent\textbf{Supplementary noise data for $\nu=4$ and $T_0=22.5~$mK.}}
We show in supplementary Fig.~S8 the noise data used to extract the electronic heat current displayed in article Fig.~3B for $\nu=4$ and $T_0=22.5~$mK. The fit of the $n=4$ reference configuration (continuous line) yields the value $\Sigma\Omega=6.5~\mathrm{nW}/\mathrm{K}^5$ for the electron-phonon coupling and $\alpha_4=4.2$ for the electronic heat flow. The data for $n=3$ is an average over the two identical configurations $(n_1=1,n_2=2)$ and $(2,1)$; similarly, the $n=5$  data is an average over the two identical configurations $(3,2)$ and $(2,3)$.\\

{\noindent\textbf{Supplementary noise data at $T_0=40~$mK.}}
We here demonstrate the robustness of our result with respect to the base temperature $T_0$.
Supplementary Fig.~S9 displays the data obtained at filling factor $\nu=3$ for a higher temperature $T_0=40~$mK. The range of injected power was adjusted to fulfill the criteria of dominant electronic heat flow in the reference configuration $n=4$ ($n\Je \gtrsim \Jeph$ at $n=4$). We used the same analysis as the one presented in the main text, the fit of the reference $n=4$ signal (continuous line in supplementary Fig.~S9A) yielding $\Sigma\Omega=5.5~\mathrm{nW}/\mathrm{K}^5$ for the electron-phonon coupling and $\alpha_4=3.8$ for the electronic heat flow. Our extraction of the electronic heat currents (supplementary Fig.~S9B) is in reasonable agreement with the expected value $(n-4) \times \frac{\pi^2 k_B^2}{6h}$ displayed as continuous lines for $n$ running from two to four. The `model-free' analysis detailed in the article yields for $\alpha_{n-4}'$, the value $-2.2 \frac{\pi^2 k_B^2}{6h}$ at $n=2$ and  $-1.2 \frac{\pi^2 k_B^2}{6h}$ at $n=3$.\\